# Psychophysiological Correlations with Gameplay Experience Dimensions

**Anders Drachen**

IT University of Copenhagen

Rued Langgaards Vej 7

2300 Copenhagen S, Denmark

drachen@itu.dk

**Lennart E. Nacke**

Blekinge Institute of Technology

Biblioteksgatan 4

374 35 Karlshamn, Sweden

Lennart.Nacke@acm.org

**Georgios Yannakakis**

IT University of Copenhagen

Rued Langgaards Vej 7

2300 Copenhagen S, Denmark

Yannakakis@itu.dk

**Anja L. Pedersen**

Userneeds A/S

Gammel Kongevej 120

Frederiksberg C 1850, Denmark

## Abstract

In this paper, we report a case study using two easy-to-deploy psychophysiological measures – electrodermal activity (EDA) and heart rate (HR) – and correlating them with a gameplay experience questionnaire (GEQ) in an attempt to establish this mixed-methods approach for rapid application in a commercial game development context. Results indicate that there is a statistically significant correlation ($p < 0.01$) between measures of psychophysiological arousal (HR, EDA) and self-reported UX in games (GEQ), with some variation between the EDA and HR measures. Results are consistent across three major commercial First-Person Shooter (FPS) games.

## ACM Classification Keywords

K.8.0 [General]: Games – Personal Computing; J.4 [Computer Applications]: Sociology, Psychology – Social and Behavioral Sciences.

## Introduction

Digital games are designed to entertain. Therefore the emotional aspects of user experience (UX) when interacting with games is of direct interest to game developers. Emotions in games act as a motivator for the cog-nitive decisions players make during gameplay [4]. Many methodologies exist for evaluating UX in digital games, which in this context is commonly referred to as player experience (PX). These can generally be divided into qualitative/semi-quantitative and quantitative methods [5]. The latter is mainly formed by psychophysiological measures [7]; controlled measures of emotion [3], generally utilized in a laboratory context [4,5]. Psychophysiological methods investigate the relationship between psychological manipulation and the resulting physiological activity [7]. Psychophysiological (i.e., signal acquisition from the human body) and psychological or sociological (i.e., questionnaires) methods are usually applied in tandem to increase the reliability of measurements. Heart rate (HR) and electrodermal activity (EDA) form two measures of activity of the peripheral nervous system, which have been successfully used as a measure of arousal and emotion [3,4,7]. While qualitative methods are often applied jointly with psychophysiological methods to increase the result reliability, there is limited knowledge available about the correlation of qualitative and quantitative player experience measures. Empirical research addressing this issue is important to facilitate maturation and growth of mixed-methods approaches. Those transport the advantages of quantitative as well as qualitative methods to game industry user researchers. Given the resource constraints of industry-based PX testing, there is also a need for exploring the potential of easy-to-deploy psychophysiological measures. In this position paper, experimental results are reported that address this issue for three different major commercial First-Person Shooter (FPS) games (Prey, Doom 3, BioShock). Using those games as experimental stimuli, we correlated PX questionnaire responses with HR and EDA; some of the easiest-to-deploy psychophysiological measures, making them potentially applicable in an industrial game development context, where time and resources can prevent the use of more cumbersome approaches (e.g., EMG, EEG or fMRI). FPS games are ideally suited to accompany measures of arousal because they are high-intensity games.

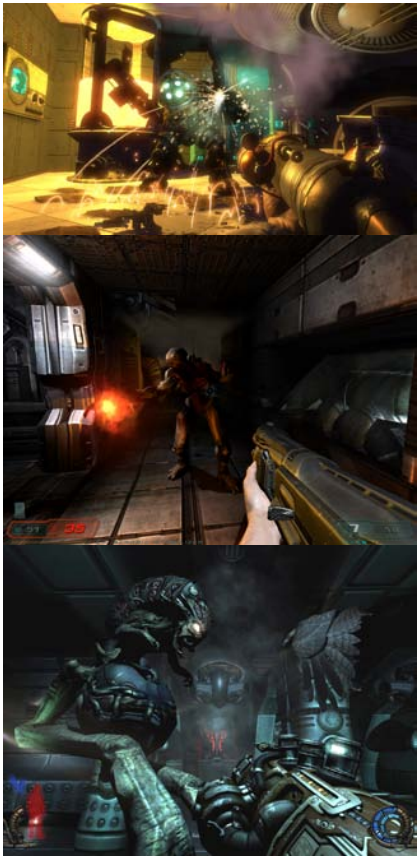

**Figure 1.** Screenshots from the three games used in the experiment. (top) Prey – developed and published by Human Head Studios, Inc. (2006). (middle) Doom 3 – developed by Id Software, Inc., and published by Activision Publishing, Inc. (2004). (bottom) Bioshock – developed and published by 2K games (2007). All images © the respective publishing companies.

## Approach and method

*Self-report Data*
The In-Game Experience Questionnaire (iGEQ) is a self-report scale for the exploration of the participant's experience during playing a computer game [1]. The survey has been tested multiple times under experimental conditions involving psychophysiological measures [4,5]. It contains 14-items, all rated on a Likert-scale scored from 0-4, distributed in pairs between seven (GEQ) dimensions of player experience: 1) Immersion (sensory and imaginative), 2) Flow, 3) Competence, 4) Tension, 5) Challenge, 6) Negative affect and 7) Positive affect.

*Heart Rate*
The cardiovascular system contains organs regulating blood flow in the body, with the heart forming a key component. Heart rate (HR) has been successfully used as a indicator for arousal [3]. A Garmin Forerunner 50 sport watch with HR monitor was used in the experiment. The device measures HR via a wireless HR monitor strapped to the participant's chest.

*Electrodermal Activity*
This is a measure of the electrical conductivity of the skin. Sweat is produced from the *eccrine* sweat glands when the central nervous system is activated (i.e., when people experience physical arousal). The sweat changes the conductivity of the skin. Research using picture stimuli has shown EDA to be correlated with self-reported emotional arousal [2], reflecting emotional responses and cognitive activity. A Thought Technologies ProComp Infinity bio-sensor system was used to measure EDA. The system measures EDA through two sensors attached to the ring and little finger on the mouse-using hand (these fingers are not used during mouse operation and least intrusive).

Three FPSs were used in the experiment (*Doom 3*, *Prey* and *BioShock*). In a FPS game, the player controls a single character. For each game a level was selected early in the game storyline. All games were set on a medium/default difficulty and played in single player mode.
*Test Procedure:* A repeated-measures experiment design was utilized. There were 3 levels of the independent variable (Prey, Doom 3 and Bioshock) and 3 dependent variables (EDA, HR, iGEQ). The psychophysiological measures were ratio scale, self-report data interval scale, permitting Pearson's correlation coefficient to be used to evaluate covariance. The experiment was run with one participant at a time, who begun by completing a consent form. Demographic data was recorded (4 participants were novice gamers, 6 intermediate and 6 hardcore gamers), and sensors fitted and tested. Before each game session, one minute baselines were recorded to permit normalization. The playing order was randomized between participants. Participants played each game for 20 minutes, following an introduction to the game controls if needed. After each 20 minute period, the participant was interviewed about their perceived PX. Participants were given as much time as needed to familiarize themselves with the game and controls. Within each playing period, every 5 minutes the game was paused by the participant, who then completed the 14-item iGEQ. This resulted in 192 completed iGEQs, 64 per game. Completing the survey generally took less than one minute and was reported as minimally interruptive to PX.

| GEQ Dimensions | HR | EDA |
|---|---|---|
| Competence | -0.36 | -0.08 |
| Immersion | -0.43 | -0.23 |
| Flow | -0.25 | -0.24 |
| Positive Affect | -0.42 | -0.20 |
| Challenge | -0.31 | -0.18 |
| Tension | 0.37 | 0.02 |
| Negative Affect | 0.24 | 0.38 |

**Table 1.** Columns *HR* and *EDA* show Pearson's correlation coefficient with *GEQ* dimensions [1]. For *HR*, significance was at least $p < 0.01$ for all *GEQ* dimensions, while *EDA* only was significant with the *GEQ* "Negative Affect" dimension (difference in dataset size means difference in *r*-critical values). *HR* was recorded as *beats per minute* (measured in 5 sec. intervals). *EDA* was continuously measured in $\mu S$.

*Data Processing*
Denoising of the psychophysiological data was performed using a modified version of the Matlab program of Yannakakis et al. [8]. Denoising and feature extraction is a multi-stage process, with initial noise reduction using discrete Fourier transformation and a moving average. The resulting signal is used for feature extraction. 95% of the HR-data were of a good enough quality to use for further analysis. For the EDA-data, a faulty sensor rendered 31% of the samples unusable, leaving 69% of the data following the de-noising and feature extraction process. The psychophysiological measurements were subsequently denoised and normalized after the average of the three recorded baselines (a *tonic* approach to measurement). By normalizing HR and EDA, the game-related effects are isolated. Each 5-min. segment came with an iGEQ per participant. Pearson's r was calculated for iGEQ values and psychophysiological measures across the three games (single-feature analysis).

## Correlation Results
The results of the correlation analysis indicate that there are statistically significant correlations between HR/EDA and GEQ dimensions; however, with different patterns of covariance (see Table 1).

*Heart Rate*
HR correlates indicate that in the three FPS games, a high HR is indicative of the participant psychologically feeling frustrated and tense. Conversely, a low HR indicates a feeling of positive affect, being in a Flow state, feeling competent and immersed in the game, despite low levels of challenge. In addition, the lower the minimal HR, maximal HR and average HR (additional features extracted using the process described above) of the participant during a 5-minute period of measurement, the higher the reported feelings of the positive GEQ dimensions (Immersion, Competence, Flow, Challenge and Positive Affect). Conversely, higher scores correlate with higher reported Tension and Negative Affect. These results indicate that a high $HR_{avg}$ value is indicative of player frustration, which can explain the covariation with the Tension and Negative Affect dimensions of the iGEQ. This result is in contrast to the findings of Mandryk and Atkins [3], who reported no correlation between high or low HR and frustration. In the study, an ice hockey game was used with two participants competing against each other. In the current experiment, fast-paced FPS games were used, although the FPSs were played in single-player mode. This provides a measure of control of the pace of the game to the player, which could be the explanation for the difference in the results of the current experiment and those from Mandryk et al. [3]. If this explanation is correct, it indicates that different games may have dif-

*Anders Drachen's Biography*

Dr. Anders Drachen is a post-doctoral research fellow at the Center for Computer Games Research at the IT University of Copenhagen. He has worked in the field of user-experience testing in games since 2002, focusing on applying a strong empirical basis to the development of practical methodologies for analyzing user behavior and user experience in computer games, and has published more than three dozen publications on these topics as well as given several invited talks.

*Lennart Nacke's Biography*

Lennart Nacke is a Ph.D. candidate in Digital Game Development at Blekinge Institute of Technology. As much as an avid gamer, he is a passionate scientist, whose research interests are psychophysiological player testing for example with EEG (i.e., brainwaves) and EMG (i.e., facial muscle contractions) as well as quantification of gameplay experience in player-game interaction and innovative interaction design with digital entertainment technologies.

ferent relationships between psychophysiological signals and self-reported UX. Since the results provided here are consistent across three different FPS, it might be assumed that correlates between HR and self-reported UX-dimensions are consistent in game genres/types.

*Electrodermal activity*

The correlation analysis indicates that EDA correlates with Negative Affect (possibly frustration). However, there is no significant correlation with the Challenge dimension of the iGEQ-survey. This is also in disparity to the findings of Mandryk et al. [3], who noted that high levels of arousal can be indicative of a high level of challenge, frustration, and/or excitement. The explanation for the difference in the results between the two studies could be related to the phrasing of the two items in the iGEQ-survey which comprise the items measuring the Challenge dimension: *I felt challenged* and: *I felt stimulated* [1]. The first relates to the feeling of challenge; however, the second refers to a positive feeling of stimulation, which does not necessarily match the definition of challenge by Mandryk et al. [3]. The two items covariate significantly, indicating that the Challenge dimension, as defined by the iGEQ survey, could have the opposite relation to EDA as Challenge defined by Mandryk et al. [3].

## Conclusion

The results indicate that HR, as a measure of arousal, is a good correlator with self-report measures of PX, either positively or negatively. Conversely, EDA is only a strong (positive) correlator with the Negative Affect dimension of the iGEQ. The difference may be related to the one-to-many relation between psychological processing and physiological responses [2], which allow specific psychophysiological measures to be linked with different psychological constructs. This highlights the importance of empirical research investigating the correlation of patterns of physiological measurement with subjective characterizations of emotions and UX [4]. The results lend support to the continued work on developing easy-to-deploy mixed-method approaches to UX-research in a game development and interactive entertainment context.